\documentclass{article}

\usepackage{cite}
\usepackage{amssymb}
\usepackage{amsmath}
\usepackage{amsfonts}
\usepackage{amsthm}
\usepackage[all]{xy}
\usepackage{enumerate}
\usepackage{colonequals}
\usepackage{mathrsfs} 
\usepackage[dvipsnames]{xcolor}
\usepackage{graphicx}
\usepackage{enumitem}
 \usepackage{mathtools}
 \usepackage{tikz}
\usepackage{multicol}

\usepackage{multicol}

\usepackage{url}

\usepackage[mathscr]{euscript}

 \usepackage{relsize}

\newtheorem{theorem}{Theorem}[section]
\newtheorem{proposition}{Proposition}[section]

\newtheorem*{theorem*}{Theorem}
\newtheorem{corollary}{Corollary}[section]
\newtheorem{definition}{Definition}[section]

\newtheorem{lemma}{Lemma}[section]

\newtheorem{remark}{Remark}[section]

 \newcommand{\freeprod}{\Huge \mathlarger{\mathlarger{*}} \normalsize }

\numberwithin{equation}{section}

\allowdisplaybreaks[2]

\newcommand{\deee}{\hspace{2 pt} \mathrm{d}}

\newcommand{\ov}[1]{\overline{#1}}

\renewcommand{\hat}{\widehat}

\allowdisplaybreaks

\begin{document}

\title{Statistical constructions in quantum information theory}
\author{Peter Burton}

\maketitle

\begin{abstract} We introduce a notion of strategies based on averaging for nonlocal games in quantum information theory. These so-called statistical strategies come in a commuting type and a more specific spatial type, which are respectively special cases of the quantum commuting and quantum spatial strategies commonly considered in the field. We prove a theorem that the sets of statistical commuting strategies and statistical spatial strategies are respectively equal to the sets of quantum commuting strategies and quantum spatial strategies for any nonlocal game. Thus we are able to use the recent negative solution of Tsirelson's problem in \cite{2020arXiv200104383J} to obtain a statistical analog showing that there exists a nonlocal game where the set of statistical commuting strategies properly contains the closure of the set of statistical spatial strategies. The proof of this theorem involves development of statistical replicas for numerous constructions in quantum information theory, in particular for the Fourier-type duality between observation structures and dynamical structures. The main point of the argument is to apply the established theory of approximating unitary representations of countable discrete groups by ergodic measure preserving actions of such groups. We note that the relevant groups are nonamenable. We also give an explicit description of a statistical strategy to win the CHSH game from Aspect's experiment with a probability exceeding the maximum possible value for a classical strategy.   \end{abstract}

\tableofcontents

\section{Introduction}

\subsection{Generalities on quantum information theory}

A nonlocal game is a type of quantum system designed to test concepts of entanglement. More specifically, in a nonlocal game two parties at physical distance from each other make observations of particles which previously interacted but have since travelled to their separate locations. The Einstein-Podolsky-Rosen hypothesis presented in \cite{PhysRev.47.777} suggested that in such a situation the observers' measurements can always be explained in terms of sampling from random variables localized to their distinct positions. In \cite{bell1964einstein}, Bell showed that this kind of local hidden variable theory places numerical restrictions on the possible correlations between the parties' measurements. Known as Bell inequalities, these restrictions have been violated in numerous experiments starting from that of Aspect in \cite{PhysRevLett.49.91} and continuing to recent 'loophope closing' results such as \cite{PhysRevLett.115.250401}, \cite{hensen2015loophole} and \cite{PhysRevLett.115.250402}. These counterexamples to local hidden variable theories can be understood as empirical evidence that the particles are genuinely entangled in the sense that their states are not determined by any physically separated quantities. \\
\\
Mathematically, a nonlocal game is thought of the process of maximizing a linear functional over certain kinds of convex subsets of Euclidean space. The convex sets over which the functional is maximized are defined as all points which can be constructed as a specific type of configuration of diagonal matrix coefficients of projection valued measures on Hilbert spaces. Well known examples of nonlocal games include the CHSH game from \cite{PhysRevLett.23.880} which we discuss in detail in Section \ref{sec.CHSH}, the GHZ three-player game as exposited in \cite{2007arXiv0712.0921G} and the Mermin-Peres magic square games described in \cite{aravind2004quantum}. For a general reference on nonlocal games we refer the reader to \cite{scarani2019bell}.\\
\\
The projection valued measures in a nonlocal game come in the structure of a multipartite graph where the pairs connected by edges commute and the relevant configuration of their matrix coefficients is referred to as a quantum commuting strategy for the nonlocal game. When the commutativity between projection valued measures is imposed by placing the varies of the multipartite graph in different factors of tensor product we obtain a quantum spatial strategy for the nonlocal game. For many years, there was a famous unsolved Tsirelson's problem (introduced in \cite{MR1264048}) which asserted that the closure of the set of quantum spatial strategies saturates the entire set of quantum commuting strategies for any nonlocal game. In \cite{fritz2012tsirelson} it was shown that Tsirelson's problem is equivalent to Connes' embedding conjecture in operator algebras. Relevant work studying issues related to Tsirelson's problem includes \cite{MR3916992}, \cite{MR4091492},  \cite{MR3898717} and \cite{MR4066471}. \\
\\
A recent breakthrough in the theory of quantum computation known as $\mathsf{MIP}^\ast = \mathsf{RE}$ from \cite{2020arXiv200104383J} resolved Tsirelson's problem in the negative and thereby refuted Connes' embedding conjecture. An interesting consequence of the negative solution to Tsirelson's problem is that it poses a physical problem to interpret the distinction between commuting and spatial strategies, as previously both types were interpreted as representing distance between the players.

\subsection{From quantum to statistical}

The purpose of this paper is to start with the specialization from a `quantum' context based on Hilbert spaces to a `statistical' context based on probability spaces and carry it forward to obtain a statistical analog of the negative solution to Tsirelson's problem. The main idea in adapting the theory of nonlocal games is to restrict to projection valued measures which can be obtained as a sequence of differences of averages. This first part of this dictionary from quantum objects to statistical objects is developed in Section \ref{sec.dictionary}. This line of reasoning allows us to define a concept of statistical strategies for nonlocal games, which appears as Definition \ref{def.measstrat} in Section \ref{sec.maindef}. We can find an analog of the distinction between quantum commuting and quantum spatial strategies in the distinction between a local product structure and a global product structure in the statistical observations. \\
\\
Our main result Theorem \ref{thm.main} in Section \ref{sec.mainthm} is to show that each the two kinds of statistical strategies define exactly the same convex sets as than their quantum counterparts. From this we are able to deduce that separation between quantum commuting and quantum spatial strategies for a particular nonlocal game entails separation between the statistical commuting and spatial strategies for the same nonlocal game. We can think of this as resolving the statistical version of Tsirelson's problem in the negative. This again poses a physical problem to interpret the difference between the two models beyond the common idea of distance between the players.\\
\\
The proof of Theorem \ref{thm.main} in Section \ref{sec.main} involves an extensive development of statistical replicas for objects in quantum information theory, most notably replicating the Fourier transform duality between projection valued measures and unitary representations of finite cyclic groups. The statistical side of this duality connects averaging procedures with measure preserving actions of finite cyclic groups. The dual object to the entire framework of projection valued measures producing one player's quantum strategy for the nonlocal game is a unitary representation of a certain nonamenable discrete group produced as a direct product of free products of the finite cyclic groups. The dual object to the entire framework of observation procedures producing one player's strategy for the nonlocal game is an ergodic measure preserving action of the same group. The distinction between commuting strategies and spatial strategies in this context translates to the distinction between commuting representations or actions and tensor/direct products of representations/actions. The theory of approximating unitary representations of countable discrete groups by ergodic measure preserving actions of such groups is well developed (see \cite{MR2583950}) and once we have built our theory to connect with that context Theorem \ref{thm.main} follows from standard constructions.\\
\\
In order to illustrate the ideas of the definitions in Section \ref{sec.def} more concretely, before proving Theorem \ref{thm.main} in Section \ref{sec.main} we give an elementary construction of an entangled statistical spatial strategy for the CHSH game in Section \ref{secsec.CHSH}. This game is a mathematical model for Aspect's experiment from \cite{PhysRevLett.49.91} that was the original situation where quantum advantage in a nonlocal game was verified empirically.

\subsection{Notation}

\begin{itemize} \item For $n \in \mathbb{N}$ write $\mathbb{Z}_n$ for the additive group $\mathbb{Z}/n\mathbb{Z}$ and $\ell^2(n)$ for $\ell^2(\{1,\ldots,n\})$. \item For $x \in \mathbb{R}$ write $e(x) = e^{2 \pi i x}$ and $e_n(x) = e\left(\frac{x}{n}\right)$. \item We assume all Hilbert spaces are separable. We assume Hilbert spaces have complex scalars except when stated otherwise. If $H$ is a complex Hilbert space we write $\mathrm{U}(H)$ for the unitary group of $H$. \end{itemize}

\subsection{Acknowledgements} We thank Lewis Bowen for suggesting that we investigate this topic and Thomas Vidick for suggesting the example of the CHSH game. We also thank the anonymous referee for helpful suggestions which improved the readability of the paper.

\section{Strategies in nonlocal games}  \label{sec.def}

\subsection{Review of nonlocal games} \label{sec.x-nonlocal}

Below are the basic definitions for the part of quantum information theory relevant to this paper.

\begin{definition} We define a \textbf{nonlocal game} to consist of the following data. \begin{itemize} \item Finite \textbf{question sets} denoted $\mathscr{X}$ and $\mathscr{Y}$ \item Finite \textbf{answer sets} denoted $\mathscr{A}$ and $\mathscr{B}$ \item A probability measure $\pi$ on $\mathscr{X} \times \mathscr{Y}$ \item A \textbf{payoff function} $D: \mathscr{X} \times \mathscr{Y} \times \mathscr{A} \times \mathscr{B} \to \{0,1\}$ \end{itemize} \end{definition}

\begin{definition} Let $\mathfrak{G} = (\mathscr{X},\mathscr{Y},\mathscr{A},\mathscr{B},\pi,D)$ be a nonlocal game.  A \textbf{bare strategy} for $\mathfrak{G}$ consists of a function $p_{x,y}:\mathscr{A} \times \mathscr{B} \to [0,1]$ for each pair $(x,y) \in \mathscr{X} \times \mathscr{Y}$ satisfying \[ \sum_{a \in \mathscr{A}} \sum_{b \in \mathscr{B}} p_{x,y}(a,b) = 1 \] If $\mathbf{p}= (p_{x,y})_{(x,y) \in \mathscr{X} \times \mathscr{Y}}$ is a bare strategy, we define the \textbf{value of the game} $\mathfrak{G}$ at $\mathbf{p}$ to be the quantity \[ \mathfrak{G}(\mathbf{p}) = \sum_{x \in \mathscr{X}} \sum_{y \in \mathscr{Y}} \pi(x,y) \sum_{a \in \mathscr{A}} \sum_{b \in \mathscr{B}} D(x,y,a,b)p_{x,y}(a,b) \]   \end{definition}

The intuition behind these definitions is that the sets $\mathscr{X}$ and $\mathscr{Y}$ represent questions asked by a referee to players Alice and Bob respectively in a cryptographic game. The players collaborate in the game to maximize their payoff. In each round, a pair of questions $(x,y) \in \mathscr{X} \times \mathscr{Y}$ is asked at random according to the distribution $\pi$ and the players respond with a pair of answers $(a,b) \in \mathscr{A} \times \mathscr{B}$ where Alice chooses $a$ and Bob chooses $b$. Prior to playing the game they are given all the information specified in its definition and they develop a family of strategy distributions $p_{x,y}$ in order to respond to the question pair $(x,y)$ with an answer pair chosen at random from $\mathscr{A} \times \mathscr{B}$ according to $p_{x,y}$. For each such round, they receive a payoff of $D(x,y,a,b)$. The value of the game at a strategy is the expected payoff from using that strategy. Note that the value of a nonlocal game at any bare strategy is between $0$ and $1$.\\
\\
Typically some restrictions are placed on bare strategies that represent physical assumptions about the relationship between the players. We now introduce the standard types of such restrictions imposed in quantum information theory. Recall that a projection valued measure on a Hilbert space is a family of orthogonal projections $A_1,\ldots,A_n$ on $H$ such that $A_jA_k = 0$ for all distinct $j,k \in \{1,\ldots,n\}$ and such that $A_1+\cdots+A_n$ is the identity operator on $H$. 

 \begin{definition} Let $\mathfrak{G} = (\mathscr{X},\mathscr{Y},\mathscr{A},\mathscr{B},\pi,D)$ be a nonlocal game. \begin{itemize} \item We define a bare strategy $(p_{x,y})_{(x,y) \in \mathscr{X} \times \mathscr{Y}}$ for $\mathfrak{G}$ to be \textbf{quantum commuting strategy} if it is generated as follows. Consider a Hilbert space $L$ and assume that for each $x \in \mathscr{X}$ we have a projection valued measure $(A^x_a)_{a \in \mathscr{A}}$ on $L$ belonging to Alice and for each $y \in \mathscr{Y}$ a projection valued measure $(B^y_b)_{b \in \mathscr{B}}$ on $L$ belonging to Bob. We assume these satisfy $A^x_aB^y_b = B^y_bA^x_a$ for all $(x,y,a,b) \in \mathscr{X} \times \mathscr{Y} \times \mathscr{A} \times \mathscr{B}$. Then we set \begin{equation} \label{eq.qucom-1} p_{x,y}(a,b) = \langle A^x_a \psi,\, B^y_b \psi  \rangle \end{equation} for some unit vector $\psi \in L$ called a \textbf{wavefunction}. We define the \textbf{quantum commuting strategy space} of $\mathfrak{G}$ by \[ \mathsf{QCo}(\mathfrak{G}) =\Bigl  \{ \mathbf{p} \in [0,1]^{\mathscr{X} \times \mathscr{Y} \times \mathscr{A} \times \mathscr{B}} : \mathbf{p} \mbox{ is a quantum commuting strategy for }\mathfrak{G} \Bigr \}  \] and define the \textbf{quantum commuting value} of $\mathfrak{G}$ to be \[ \mathrm{val}^{\mathsf{Co}}(\mathfrak{G}) = \sup_{\mathbf{p} \in \mathsf{QCo}(\mathfrak{G})} \mathfrak{G}(\mathbf{p})\]   \item We define a quantum commuting strategy $(p_{x,y})_{(x,y) \in \mathscr{X} \times \mathscr{Y}}$ for $\mathfrak{G}$ to be a \textbf{quantum spatial strategy} if it is generated as follows. Consider Hilbert spaces $H$ and $K$ and assume for each $(x,y) \in \mathscr{X} \times \mathscr{Y}$ we have a projection valued measure $(A_x^a)_{a \in \mathscr{A}}$ on $H$ belonging to Alice and a projection valued measure $(B_y^b)_{b \in \mathscr{B}}$ on $K$ belonging to Bob. Then we set \begin{equation} \label{eq.spatial} p_{x,y}(a,b) = \langle  (A_x^a \otimes B_y^b) \psi,\, \psi \rangle \end{equation} for some unit vector $\psi \in H \otimes K$. We define the \textbf{quantum spatial strategy space} of $\mathfrak{G}$ by \[ \mathsf{QSp}(\mathfrak{G}) =\Bigl  \{ \mathbf{p} \in [0,1]^{\mathscr{X} \times \mathscr{Y} \times \mathscr{A} \times \mathscr{B}} : \mathbf{p} \mbox{ is a quantum spatial strategy for }\mathfrak{G} \Bigr \}  \] and define the \textbf{quantum spatial value} of $\mathfrak{G}$ to be \[ \mathrm{val}^\ast(\mathfrak{G}) = \sup_{\mathbf{p} \in \mathsf{QSp}(\mathfrak{G})} \mathfrak{G}(\mathbf{p})\] \item We define a quantum spatial strategy $(p_{x,y})_{(x,y) \in \mathscr{X} \times \mathscr{Y}}$ to be a \textbf{classical strategy} if there exists $\varsigma \in \ell^1(\mathbb{N})$ such that $||\varsigma||_1 = 1$ and \begin{equation} \label{eq.entangle-1} \psi  = \sum_{j=1}^\infty \varsigma_j (\phi_j \otimes \omega_j )\end{equation} for $\phi_j \in H$ and $\omega_j \in K$ with $||\phi_j|| = ||\omega_j|| = 1$. We define the \textbf{classical value} of $\mathfrak{G}$ to be \[ \mathrm{val}^{\mathsf{Cl}}(\mathfrak{G}) = \sup \Bigl \{ \mathfrak{G}(\mathbf{p}): \mathbf{p} \mbox{ is a classical strategy for }\mathfrak{G} \Bigr \} \] \end{itemize} \label{def.quantum} \end{definition}
 
The commutativity and tensor product hypotheses represent spatial separation between the players. The players use their projection valued measures to collapse the wavefunction $\psi$ of a particle and thereby sample from the distributions comprising their strategy. If $\mathfrak{G}(\mathbf{p}) > \mathrm{val}^{\mathsf{Cl}}(\mathfrak{G})$ then the strategy $\mathbf{p}$ is referred to as entangled and the players are understood to be achieving some measure of quantum mechanical coordination in the game that is impossible when $\psi$ is localized to each player as in a classical strategy. More explicitly, this means that a decomposition such as (\ref{eq.entangle-1}) can achieved in general only with $||\varsigma||_2 = 1$ and $||\varsigma||_1$ going to infinity. \\
\\
We note that we can regard $\mathfrak{G}$ as discrete information in the context of computability theory. The following is proved in \cite{MR2513501} and \cite{Navascu_s_2008}.

\begin{theorem} \label{thm.inner} There exists a procedure which takes $\mathfrak{G}$ and computes a function $u_\mathfrak{G}:\mathbb{N} \to [0,1]$ such that the sequence $u_\mathfrak{G}(n) - \mathrm{val}^{\mathsf{Co}}(\mathfrak{G})$ is nonnegative and converges to zero. \end{theorem}

The main result of \cite{2020arXiv200104383J} establishing $\mathsf{MIP}^* = \mathsf{RE}$ is the following.

\begin{theorem} \label{thm.MIP} In order to solve the halting problem, it suffices to have access to a procedure which takes $\mathfrak{G}$ and computes an approximation function $v_\mathfrak{G}:\mathbb{N} \to [0,1]$ and an error function $\epsilon_\mathfrak{G}:\mathbb{N} \to [0,1]$ such that $\epsilon_\mathfrak{G}$ is nonincreasing and converges to zero and $|v_\mathfrak{G}(n) - \mathrm{val}^\ast(\mathfrak{G})| \leq \epsilon_\mathfrak{G}(n)$. \end{theorem}

An exhaustive search through quantum spatial strategies makes it easy to use $\mathfrak{G}$ to compute a function $\ell_\mathfrak{G}:\mathbb{N} \to [0,1]$ such that the sequence $\mathrm{val}^\ast(\mathfrak{G})- \ell_\mathfrak{G}(n)$ is nonnegative and converges to zero. Since we can take $v_\mathfrak{G}(n) = \frac{1}{2}(u_\mathfrak{G}(n)+\ell_\mathfrak{G}(n))$ and $\epsilon_\mathfrak{G}(n) = u_\mathfrak{G}(n) - \ell_\mathfrak{G}(n)$ for a game where $\mathrm{val}^\mathsf{Co}(\mathfrak{G}) = \mathrm{val}^\ast(\mathfrak{G})$, from Theorems \ref{thm.inner} and \ref{thm.MIP} we deduce the following.

\begin{corollary} \label{thm.separate-1} There exists a game for which $\mathrm{val}^{\mathsf{Co}}(\mathfrak{G}) > \mathrm{val}^\ast(\mathfrak{G})$. For this game we must necessarily have that the closure of $\mathsf{QSp}(\mathfrak{G})$ is a proper subset of $\mathsf{QCo}(\mathfrak{G})$. \end{corollary}

The theory of ultraproducts of representations makes it clear that $\mathsf{QCo}(\mathfrak{G})$ is closed for every nonlocal game $\mathfrak{G}$, while in \cite{MR3898717} it was proved that there exists a game $\mathfrak{G}$ such that $\mathsf{QSp}(\mathfrak{G})$ is not closed. However, the deeper issue is whether the closure of $\mathsf{QSp}(\mathfrak{G})$ is equal to $\mathsf{QCo}(\mathfrak{G})$ for every nonlocal game $\mathfrak{G}$. Prior to the establishment of Theorem \ref{thm.MIP} this question was known as Tsirelson's problem. It was well known that verifying global equality in Tsirelson's problem was equivalent verifying to Connes' embedding conjecture, and therefore Corollary \ref{thm.separate-1} refutes Connes' embedding conjecture. The purpose of this paper is to translate these ideas into the language of probability theory.

\subsection{Statistical strategies} \label{subsec.measstrat}

\subsubsection{Probabalistic cases of linear objects} \label{sec.dictionary}

In Section \ref{sec.dictionary} we introduce a number of constructions in probability theory analogous to constructions in  Hilbert spaces. These culminate in Section \ref{sec.maindef} where we introduce the statistical analog of quantum strategies for nonlocal games.

\begin{definition} \label{def-30} For $n \in \mathbb{N}$ we define the \textbf{averaging operator} on $\ell^2(n)$ by \[ \mathbb{A}_n[f](k) = \frac{1}{n} \sum_{j=1}^n f(j) \]  for $f:\{1,\ldots,n\} \to \mathbb{C}$ and all $k \in \{1,\ldots,n\}$. Thus $\mathbb{A}_n[f]$ is constant for any $f$. We also define the identity operator $\mathbb{I}_n$ on $\ell^2(n)$. If $k \in \{0,\ldots,n\}$ we define the \textbf{partial averaging operator} $\mathbb{I}_k \oplus \mathbb{A}_{n-k}$ according to the natural decomposition $\ell^2(n) = \ell^2(k)\oplus \ell^2(n-k)$. \end{definition}

We recall that a binary relation $\sim$ on a standard probability space $(\Omega,\mu)$ is said to be measurable if the defining set $\{(s,t) \in \Omega \times \Omega: s \sim t\}$ is a measurable subset of the product measure space $\Omega \times \Omega$. If $\sim$ is an equivalence relation then a function on $\Omega$ to said to be class-bijective relative to $\sim$ if it restricts to a bijection of each of the equivalence classes of $\sim$.

\begin{definition} \label{def-31} Let $(\Omega,\mu)$ be a standard probability space referred to as the \textbf{sample space} and $n \in \mathbb{N}$. We define an \textbf{observable of resolution} $n$ on $(\Omega,\mu)$ to consist of a measurable equivalence relation $\sim_\alpha$ on $\Omega$ such that all classes have size $n$ and a measurable function $c_\alpha: \Omega \to \{1,\ldots,n\}$ which is class-bijective relative to $\sim_\alpha$. Given an observable $\alpha$, and $k \in \{0,\ldots,n\}$ we denote the \textbf{observation operator of order }$k$\textbf{ associated with }$\alpha$ on $L^2(\Omega,\mu)$ by $\mathbb{O}_{\alpha,k}$ and define it by stipulating that it restricts to the partial averaging operator $\mathbb{I}_k \oplus \mathbb{A}_{n-k}$ on each $\sim_\alpha$ class, where the identification between $\sim_\alpha$-classes and $\{1,\ldots,n\}$ is given by $c_\alpha$. \end{definition} 

The existence of $c_\alpha$ allows us to use terms like first, last and $k^{\mathrm{th}}$ for $k \in \{1,\ldots,n\}$ relative to $\sim_\alpha$ classes. We have that $\mathbb{O}_{\alpha,k}$ is the orthogonal projection from $L^2(\Omega,\mu)$ onto the subspace of functions which are constant on the last $n-k$ points in their $\sim_\alpha$-class. It is also the conditional expectation on the $\sigma$-algebra generated by a single-element cell for each point among the first $k$ of its $\sim_\alpha$-class and a single $n-k$ element cell at the end of each $\sim_\alpha$-class.

\begin{definition} Let $\alpha$ and $\beta$ be observables on $(\Omega,\mu)$ of resolution $n$ and $m$ respectively. We say that $\alpha$ and $\beta$ are \textbf{consistent} if the following objects exist. \begin{itemize} \item A measurable equivalence relation $\sim_{\alpha \times \beta}$ on $\Omega$ such that each $\sim_{\alpha \times \beta}$-class has size $nm$ and is saturated under both $\sim_\alpha$ and $\sim_\beta$ \item A measurable function $c_{\alpha \times \beta}: \Omega \to \{1,\ldots,n\} \times \{1,\ldots,m\}$ which is class-bijective relative to $\sim_{\alpha \times \beta}$ and such that $c_{\alpha \times \beta}(u) = (c_\alpha(u),c_\beta(u))$ for all $u \in \Omega$. \end{itemize} \end{definition}

Many of the fundamental ideas in the theory of projections on Hilbert spaces have statistical analogs for observation operators described in the proposition below. 

\begin{proposition} \label{prop.dictionary} It is elementary to verify all the first three assertions below for the operators $\mathbb{I}_k \oplus \mathbb{A}_{n-k}$ and then they follow immediately for $\mathbb{O}_{\alpha,k}$ by integration over $\sim_\alpha$-classes. Let $\alpha:\Omega \to [0,1]$ be an observable.

\begin{enumerate}[label=(\roman*)] \item For all $n \in \mathbb{N}$ the operator $\mathbb{O}_{\alpha,n-1}$ is the identity operator on $L^2(\Omega,\mu)$. \item If $0 \leq j \leq k \leq n$ then the projection $\mathbb{O}_{\alpha,k}$ covers the projection $\mathbb{O}_{\alpha,j}$ and so we have that  $\mathbb{O}_{\alpha,k}-\mathbb{O}_{\alpha,j}$ is a projection. In particular this implies \[  \int_\Omega \mathbb{O}_{\alpha,k}[f](u)\ov{f(u)} \deee \mu(u) - \int_\Omega \mathbb{O}_{\alpha,j}[f](u)\ov{f(u)} \deee \mu(u) \geq 0 \] for all $f \in L^2(\Omega,\mu)$. \item From the above items it is clear that for any $n \in \mathbb{N}$ the operators \[ \{\mathbb{O}_{\alpha,k} - \mathbb{O}_{\alpha,k-1} : 0 \leq k \leq n-1\} \] are a projection valued measure on $L^2(\Omega,\mu)$ with the convention that $\mathbb{O}_{\alpha,-1} = 0$ for all $\alpha$. \item Suppose $\beta:\Omega \to [0,1]$ is an observable which is consistent with $\alpha$. Then $\mathbb{O}_{\alpha,k}$ and $\mathbb{O}_{\beta,j}$ commute for all $k \in \{0,\ldots,n-1\}$ and $j \in \{0,\ldots,m-1\}$. We may see this as follows. Identify each $\sim_{\alpha \times \beta}$ class with $\{0,\ldots,n-1\} \times \{0,\ldots,m-1\}$. For $k \in \{0,\ldots,n-1\}$ and $j \in \{0,\ldots,m-1\}$ write $\mathbf{1}_{k,j}$ for the indicator function of $\{(k,j)\}$ in $\{0,\ldots,n-1\} \times \{0,\ldots,m-1\}$. Also write $\mathbf{1}_k$ for the indicator function of $\{k\}$ in $\{1,\ldots,n\}$. Then it is clear that for all $k,\ell \in \{0,\ldots,n-1\}$ and all $j,r \in \{0,\ldots,m-1\}$ we have \[ \mathbb{O}_{\beta,j}\mathbb{O}_{\alpha,k}[\mathbf{1}_{\ell,r}] = (\mathbb{I}_k \oplus \mathbb{A}_{n-k})[\mathbf{1}_\ell] \cdot (\mathbb{I}_j \oplus \mathbb{A}_{m-j})[\mathbf{1}_r]  =  \mathbb{O}_{\alpha,k}\mathbb{O}_{\beta,j}[\mathbf{1}_{\ell,r}] \] where the product in the center of the previous display denoted $\cdot$ is numerical. By linearity the commutativity holds on all of $\ell^2(\{0,\ldots,n-1\} \times \{0,\ldots,k-1\})$.  \item Suppose there exists standard probability spaces $(\Lambda,\nu)$ and $(\Pi,\eta)$ such that $(\Omega,\mu) = (\Lambda \times \Pi,\nu \times \eta)$. Let $\alpha$ be an observable on $\Lambda$ and let $\beta$ be an observable on $\Lambda$. We may lift $\alpha$ to an observable $\alpha^\circ$ on $\Lambda \times \Pi$ by taking the Cartesian product of each $\sim_\alpha$-class with each single point in $\Pi$ to obtain $\sim_{\alpha^\circ}$ and then lifting the linear order in the only way possible. We may lift $\beta$ to an observable $\beta^\circ$ on $\Lambda \times \Pi$ in a similar way. Then $\alpha^\circ$ and $\beta^\circ$ are consistent and so $\mathbb{O}_{\alpha^\circ,k}$ and $\mathbb{O}_{\beta^\circ,j}$ commute for all $k \in \{0,\ldots,n-1\}$ and $j \in \{0,\ldots,m-1\}$. \item In order to see that observation operators do not commute in general, we introduce the following example. Let $\Omega = [0,1)$, let $\mu$ be Lebesgue measure and let $n = 3$. Define an observable $\alpha$ on $[0,1)$ by declaring $x \sim_\alpha y$ if and only if $3x = 3y \,\,(\mathrm{mod}\,\, 1)$. Define $c_\alpha(x) = \lfloor 3x\rfloor+1$. Define an observable $\beta$ on $[0,1)$ by setting $\sim_\beta$ equal to $\sim_\alpha$ and $c_\beta(x) =3- \lfloor 3x \rfloor$. Define $f:[0,1) \to \mathbb{C}$ by \[ f(x) = \begin{dcases} 1 & \mbox{ if }0 \leq x < \frac{1}{3} \\ -1 & \mbox{ if } \frac{1}{3} \leq x < \frac{2}{3} \\ 0 & \mbox{ if } \frac{2}{3} \leq x < 1 \end{dcases}  \] Then we have $\mathbb{O}_{\beta,1}[f](x) = 0$ for all $x$ so that $\mathbb{O}_{\alpha,1}[\mathbb{O}_{\beta,1}[f]](x) = 0$ for all $x$. On the other hand, we have \[ \mathbb{O}_{\alpha,1}[f](x) = \begin{dcases}  1 & \mbox{ if } 0 \leq x < \frac{1}{3}   \\ - \frac{1}{2}&\mbox{ if } \frac{1}{3} \leq k < 1 \end{dcases} \] Therefore \[ \mathbb{O}_{\beta,1}[\mathbb{O}_{\alpha,1}[f]]\left( \frac{5}{6}\right)  = -\frac{1}{2} < 0 = \mathbb{O}_{\alpha,1}[\mathbb{O}_{\beta,1}[f]]\left(\frac{5}{6}\right) \]   \end{enumerate}
\end{proposition}

 \subsubsection{Main definition} \label{sec.maindef}

We now state the main definition of this paper, which is a concept of a strategy for a nonlocal game that is based on statistical considerations. 

\begin{definition} \label{def.measstrat} Let $\mathfrak{G} = (\mathscr{X},\mathscr{Y},\mathscr{A},\mathscr{B},\pi,D)$ be a nonlocal game. Enumerate $\mathscr{A} = \{a_1,\ldots,a_n\}$ and $\mathscr{B} = \{b_1,\ldots,b_m\}$. \begin{itemize} \item We define a bare strategy for $\mathfrak{G}$ to be a \textbf{statistical commuting strategy} if it is generated by the following data. \begin{itemize} \item A standard probability space $(\Omega,\mu)$ which represents a common sample space belonging to both Alice and Bob. \item For each $x \in \mathscr{X}$ let $\alpha_x$ be an observable on $\Omega$ belonging to Alice. For $k \in \{1,\ldots,n\}$ we write $\mathbb{O}_{x,k}$ for $\mathbb{O}_{\alpha_x,k}$. \item For each $y \in \mathscr{Y}$ let $\beta_y$ be an observable  on $\Omega$ belonging to Bob. For $j \in \{1,\ldots,m\}$ write $\mathbb{O}_{y,j}$ for $\mathbb{O}_{\beta_y,j}$. \item We stipulate that for each pair $(x,y) \in \mathscr{X} \times \mathscr{Y}$ we have that $\alpha_x$ is consistent with $\beta_y$. \item A measurable function $f:\Omega \to \mathbb{C}$ with \[ \int_\Omega |f(u)|^2 \deee \mu(u) = 1 \]  called a \textbf{wavefunction}. \end{itemize} Given these data we set $p_{x,y}(a_k,b_j)$ to be the quantity  \begin{equation} \int_\Omega \bigl(\mathbb{O}_{x,k}[f](u) - \mathbb{O}_{x,k-1}[f](u)\bigr)\bigl( \overline{\mathbb{O}_{y,j}[f](u)} - \overline{\mathbb{O}_{y,j-1}[f](u)} \bigr) \deee \mu(u)   \label{eq.excom-1} \end{equation} By comparing (\ref{eq.qucom-1}) with (\ref{eq.excom-1}) in light of Items $(\mathrm{iii})$ and $(\mathrm{iv})$ of Proposition \ref{prop.dictionary} we see that statistical commuting strategies are special cases of quantum commuting strategies. We define the \textbf{statistical commuting strategy space} of $\mathfrak{G}$ by \[ \mathsf{StatCo}(\mathfrak{G}) =\Bigl  \{ \mathbf{p} \in [0,1]^{\mathscr{X} \times \mathscr{Y} \times \mathscr{A} \times \mathscr{B}} : \mathbf{p} \mbox{ is a statistical commuting strategy for }\mathfrak{G} \Bigr \}  \] and define the \textbf{statistical commuting value} of $\mathfrak{G}$ to be \[ \mathrm{val}^{\mathsf{Co}}_{\mathsf{St}}(\mathfrak{G}) = \sup_{\mathbf{p} \in \mathsf{StatCo}(\mathfrak{G})} \mathfrak{G}(\mathbf{p})\]   \item We define a statistical commuting strategy for $\mathfrak{G}$ to be a \textbf{statistical spatial strategy} if it is generated by the following data. \begin{itemize} \item Standard probability spaces $(\Lambda,\nu)$ and $(\Pi,\eta)$ which represent two spatially separated sample spaces belonging to Alice and Bob respectively.  \item For each $x \in \mathscr{X}$ an observable $\alpha_x:\Lambda \to [0,1]$ belonging to Alice. Write $\mathbb{O}^\circ_{x,k}$ for the observable on $L^2(\Lambda \times \Pi,\nu \times \eta)$ given by $\mathbb{O}_{\alpha_x^\circ,k}$ where $\alpha_x^\circ$ is as in Item $\mathrm{(v)}$ of Proposition \ref{prop.dictionary}.   \item For each $y \in \mathscr{Y}$ an observable $\beta_y:\Pi \to [0,1]$ belonging to Bob. Write $\mathbb{O}^\circ_{y,j}$ for the observable on $L^2(\Lambda \times \Pi,\nu \times \eta)$ given by $\mathbb{O}_{\beta_y^\circ,j}$. \item A measurable function $f: \Lambda \times \Pi \to \mathbb{C}$ with \[\int_{\Lambda \times \Pi} |f(s,t)|^2 \deee(\nu \times \eta)(s,t) = 1 \]  \end{itemize} Given these data, we set $p_{x,y}(a_k,b_j)$ to  \begin{equation} \int_{\Lambda \times \Pi} \bigl(\mathbb{O}^\circ_{x,k}[f](s,t) - \mathbb{O}^\circ_{x,k-1}[f](s,t)\bigr)\bigl( \overline{\mathbb{O}^\circ_{y,j}[f](s,t)} - \overline{\mathbb{O}^\circ_{y,j-1}[f](s,t)} \bigr) \deee (\nu \times \eta)(s,t)  \label{eq.spatial-2} \end{equation} By comparing (\ref{eq.spatial}) and (\ref{eq.spatial-2}) in light of Items $(\mathrm{iii})$ and $(\mathrm{v})$ of Proposition \ref{prop.dictionary} we see that statistical spatial strategies are special cases of quantum spatial strategies. We define the \textbf{statistical spatial strategy space} of $\mathfrak{G}$ by \[ \mathsf{StatSp}(\mathfrak{G}) = \Bigl \{\mathbf{p} \in [0,1]^{\mathscr{X} \times \mathscr{Y} \times \mathscr{A} \times \mathscr{B}}: \mathbf{p} \mbox{ is a statistical spatial strategy for }\mathfrak{G} \Bigr \} \] and set \[ \mathrm{val}_{\mathsf{St}}^\ast(\mathfrak{G}) = \sup_{\mathbf{p} \in \mathsf{StatSp}(\mathfrak{G})} \mathfrak{G}(\mathbf{p}) \] \item We say a statistical spatial strategy is \textbf{classical} if there exists $\varsigma \in \ell^1(\mathbb{N})$ with $||\varsigma||_1 = 1$ such that we can write  \begin{equation} \label{eq.entangle-2} f(s,t) = \sum_{j=1}^\infty \varsigma_jg_j(s)h_j(t) \end{equation} for $g_j \in L^2(\Lambda,\nu)$ and $h_j \in L^2(\Pi,\eta)$ with \[ \int_\Lambda |g_j(s)|^2 \deee \nu(s) = \int_\Pi |h_j(t)|^2 \deee \eta(t) = 1\] for all $j \in  \mathbb{N}$. \end{itemize}  \end{definition}

The statistical entanglement is represented by the fact that a decomposition such as (\ref{eq.entangle-2}) can be achieved in general only with $||\varsigma||_2 = 1$ and $||\varsigma||_1$ going to infinity.

\subsection{Main theorem and corollaries} \label{sec.mainthm}

The main result of this paper is to prove the following theorem.

\begin{theorem} \label{thm.main} For any nonlocal game $\mathfrak{G} = (\mathscr{X},\mathscr{Y},\mathscr{A},\mathscr{B},\pi,D)$ we have that $\mathsf{QSp}(\mathfrak{G}) = \mathsf{StatSp}(\mathfrak{G})$ and $\mathsf{QCo}(\mathfrak{G}) = \mathsf{StatCo}(\mathfrak{G})$.\end{theorem} 

We obtain the following corollary of Theorems \ref{thm.inner} and \ref{thm.main}.

\begin{corollary} \label{thm.innerm} There a procedure which takes $\mathfrak{G}$ and computes a function $u_\mathfrak{G}:\mathbb{N} \to [0,1]$ such that the sequence $u_\mathfrak{G}(n) - \mathrm{val}^{\mathsf{Co}}_\mathsf{St}(\mathfrak{G})$ is nonnegative and converges to zero. \end{corollary}

We obtain the following corollary of Theorems \ref{thm.MIP} and \ref{thm.main}.

\begin{corollary}\label{thm.MIPm} In order to solve the halting problem, it suffices to have access to a procedure which takes $\mathfrak{G}$ and computes an approximation function $v_\mathfrak{G}:\mathbb{N} \to [0,1]$ and an error function $\epsilon_\mathfrak{G}:\mathbb{N} \to [0,1]$ such that $\eta_\mathfrak{G}$ is nonincreasing and converges to zero and $|v_\mathfrak{G}(n) - \mathrm{val}^\ast_{\mathsf{St}}(\mathfrak{G})| \leq \epsilon_\mathfrak{G}(n)$. \end{corollary}

An exhaustive search through statistical spatial strategies makes it easy to use $\mathfrak{G}$ to compute a function $\ell_\mathfrak{G}:\mathbb{N} \to [0,1]$ such that the sequence $\mathrm{val}^\ast_\mathsf{St}(\mathfrak{G})- \ell_\mathfrak{G}(n)$ is nonnegative and converges to zero. Thus we may use the same reasoning as was used to deduce Corollary \ref{thm.separate} from Theorems \ref{thm.inner} and \ref{thm.MIP} to deduce the following further corollary of Corollaries \ref{thm.innerm} and \ref{thm.MIPm}.

\begin{corollary} \label{thm.separate} There exists a game for which $\mathrm{val}_\mathsf{St}^{\mathsf{Co}}(\mathfrak{G}) > \mathrm{val}_\mathsf{St}^\ast(\mathfrak{G})$. For this game we must necessarily have that the closure of $\mathsf{StatSp}(\mathfrak{G})$ is a proper subset of $\mathsf{StatCo}(\mathfrak{G})$. \end{corollary}

We may regard Corollary \ref{thm.separate} as a negative solution to the statistical version of Tsirelson's problem.

\section{The CHSH game} \label{secsec.CHSH}

The CHSH game $\mathfrak{C}$ was introduced mathematically in \cite{1313847}. It was previously studied in a physical context in the papers \cite{PhysRevLett.49.91} and \cite{PhysRevLett.23.880}. In Section \ref{secsec.CHSH} we construct an explicit statistical spatial strategy $\mathbf{p}$ such that $\mathfrak{C}(\mathbf{p}) - \frac{1}{16}=\mathrm{val}^{\mathrm{cl}}(\mathfrak{C})$, thereby demonstrating that statistical spatial strategies can be nontrivially entangled.

\subsection{Standard theory of CHSH}  \label{sec.CHSH}

In Section \ref{sec.CHSH} we exposit the standard theory of the CHSH game.

\subsubsection{Definitions}

\begin{definition} The \textbf{CHSH game} is a nonlocal game defined as follows. Let $\mathscr{X} = \mathscr{Y} = \mathscr{A} = \mathscr{B} = \{0,1\}$, where we regard the $0$ bit as false and the $1$ bit as true. We define $\pi$ to be the uniform probability measure on $\mathscr{X} \times \mathscr{Y}$ and define \[ D(x,y,a,b) =  \begin{cases} 1 &\mbox{ if } x\, \mathrm{AND}\,y = a\, \mathrm{XOR}\,b \\ 0 &\mbox{ if } x\, \mathrm{AND}\,y \neq a\, \mathrm{XOR} \,b \end{cases} \] \end{definition}

We may interpret this game as follows. The referee chooses two bits $x$ and $y$ uniformly at random then sends the bit $x$ to Alice and the bit $y$ to Bob. Alice and Bob respond with bits $a$ and $b$ respectively. Alice and Bob win the game if and only if $D(x,y,a,b) = 1$. More explicitly, if $x = y = 1$ then they win their bits differ and if at least one of $x$ and $y$ is zero they win if their bits agree.\\
\\
Let $\vec{\imath} = (1,0)$ and $\vec{\jmath} = (0,1)$ be the standard basis vectors in $\mathbb{R}^2$. For $\theta \in [-\pi,\pi]$ let $q_\theta$ be the orthogonal projection on $\cos(\theta)\vec{\imath} + \sin(\theta)\vec{\jmath}$ and let $\hat{q}_\theta$ be the orthogonal projection on $\sin(\theta)\vec{\imath} - \cos(\theta) \vec{\jmath}$. Thus $q_\theta + \hat{q}_\theta = I$. Let also \[ \Delta = \frac{1}{\sqrt{2}}\vec{\imath}\otimes \vec{\imath}+\frac{1}{\sqrt{2}}\vec{\jmath} \otimes \vec{\jmath} \in \mathbb{R}^2 \otimes \mathbb{R}^2 \]

This state is known as an EPR pair and can be interpreted as a maximally entangled superposition of the states $\imath \otimes \imath$ and $\jmath \otimes \jmath$. 

\begin{definition} A \textbf{angular CHSH strategy} consists of two functions $\theta,\eta:\{0,1\} \to [-\pi,\pi]$ giving rise a strategy of the form \begin{align*} p_{x,y}(0,0) & = \langle \Delta |q_{\theta(x)} \otimes q_{\eta(y)}|\Delta \rangle \\ p_{x,y}(1,0) &= \langle \Delta |\hat{q}_{\theta(x)} \otimes q_{\eta(y)}|\Delta \rangle \\  p_{x,y}(0,1) & = \langle \Delta |q_{\theta(x)} \otimes \hat{q}_{\eta(y)}|\Delta \rangle \\ p_{x,y}(1,1) & = \langle \Delta|\hat{q}_{\theta(x)} \otimes \hat{q}_{\eta(y)}| \Delta \rangle  \end{align*}  \end{definition}

\begin{proposition} \label{prop.CHSH-elementary} The following observations are elementary.  \begin{align*} ||q_\theta( \imath)||^2 &= ||\hat{q}_\theta(\jmath)||^2 = \cos^2(\theta)   \\  ||q_\theta(\jmath)||^2 & = ||\hat{q}_\theta(\imath)||^2 = \sin^2(\theta)  \\  \langle \imath|q_\theta|\jmath \rangle &=- \langle \imath|\hat{q}_\theta|\jmath\rangle = \cos(\theta)\sin(\theta)\end{align*} \end{proposition}

We also note that the identification of $\imath$ with $\imath$ across opposite sides of the tensor product is illusory and the strategy is identical if the space $\mathbb{R}^2 \otimes \mathbb{R}^2$ is replaced with $H \otimes K$ for two-dimensional Hilbert space $H$ and $K$ and $(\imath,\jmath)$ is replaced with an arbitrary orthonormal basis for each of $H$ and $K$.

\subsubsection{Computations} \label{seg.compute}

We make the following standard computation using Proposition \ref{prop.CHSH-elementary}.

\begin{align*} 2\langle (q_\theta \otimes q_\eta)\Delta, \, \Delta \rangle & = ||q_\theta(\vec{\imath})||^2 ||q_\eta(\vec{\imath})||^2+ ||q_\theta(\vec{\jmath})||^2 ||q_{\eta}(\vec{\jmath})||^2 + 2\langle \vec{\imath}|q_\theta|\vec{\jmath}\rangle \langle \vec{\imath}|q_\eta|\vec{\jmath} \rangle  \\ & = \cos^2(\theta)\cos^2(\eta)+\sin^2(\theta)\sin^2(\eta)+2\cos(\theta)\sin(\theta)\cos(\eta)\sin(\eta)  \\ & = \cos^2(\theta-\eta) \end{align*}

We make a second similar computation.

\begin{align*}2 \langle (q_\theta \otimes \hat{q}_\eta)\Delta,\,\Delta \rangle & = 2 \langle (\hat{q}_\eta \otimes q_\theta)\Delta, \,\Delta\rangle \\ & = ||q_\theta(\vec{\imath})||^2 ||\hat{q}_\eta(\vec{\imath})||^2+ ||q_{\theta}(\vec{\jmath})||^2 ||\hat{q}_{\eta}(\vec{\jmath})||^2 +2 \langle \vec{\imath}|q_\theta|\vec{\jmath}\rangle \langle \vec{\imath}|\hat{q}_{\eta}|\vec{\jmath} \rangle\\ & = \cos^2(\theta)\sin^2(\eta)+\sin^2(\theta)\cos^2(\eta) - 2\cos(\theta)\sin(\theta)\cos(\eta)\sin(\eta)\\ &= \sin^2(\eta-\theta) \end{align*}

We make a third similar computation.

\begin{align*}2 \langle (\hat{q}_\theta\otimes \hat{q}_\eta) \Delta ,\,\Delta \rangle & = ||\hat{q}_\theta(\vec{\imath})||^2 ||\hat{q}_\eta(\vec{\imath})||^2+ ||\hat{q}_\theta(\vec{\jmath})||^2 ||\hat{q}_\eta(\vec{\jmath})||^2 +2 \langle \vec{\imath}|\hat{q}_\theta|\vec{\jmath}\rangle \langle \vec{\imath}|\hat{q}_\eta|\vec{\jmath} \\ & = \sin^2(\theta)\sin^2(\eta)+\cos^2(\theta)\cos^2(\eta)+\cos(\theta)\sin(\theta)\cos(\eta)\sin(\eta) \\ & = \cos^2(\theta-\eta) \end{align*}

\subsubsection{Conclusions}

If $(x,y) \in \{(0,0),(0,1),(1,0)\}$ then Alice and Bob win if and only if they answer $(0,0)$ or $(1,1)$ so the probability of winning in this case is $\cos^2(\eta(x)-\theta(y))$. If $(x,y) = (1,1)$ then Alice and Bob win if and only if they answer $(1,0)$ or $(0,1)$ so the probability of winning in this case is $\sin^2(\eta(x)-\theta(y))$. Therefore if we write $\mathfrak{C}(\theta,\eta)$ for the value of the CHSH game at a strategy $\{\theta,\eta:\{0,1\} \to [0,2\pi]\}$ then we have \[ \mathfrak{C}(\theta,\eta)  = \frac{1}{4}\bigl( \cos^2(\eta(0)-\theta(0))+\cos^2(\eta(0)-\theta(1))+\cos^2(\theta(0)-\eta(1))+\sin^2(\theta(1)-\eta(1) \bigr) \] It is straightforward to calculate from the last display that $\mathfrak{C}(\theta,\eta) = \frac{13}{16}$ when \begin{equation} (\theta(0) , \theta(1), \eta(0), \eta(1)) = \left(0,\frac{\pi}{3},\frac{\pi}{6},-\frac{\pi}{6} \right) \label{eq.angles} \end{equation} It is also straightforward to calculate by exhaustively listing classical strategies that $\mathrm{val}^{\mathsf{Cl}}(\mathfrak{C}) = \frac{3}{4} = \frac{13}{16} - \frac{1}{16}$, so the strategy given in (\ref{eq.angles}) is nontrivially entangled.

\subsection{Entanglement in a statistical strategy} \label{sec.ergCHSH}

In Section \ref{sec.ergCHSH} we give a construction of the strategy from (\ref{eq.angles}) for CHSH as a statistical spatial strategy.

\subsubsection{Requirements of the construction} \label{sec.requirements}

Let $\kappa \in \left\{-\frac{\pi}{6},0\right\}$ and let $\lambda = \kappa+ \frac{\pi}{3}$. Let $\Omega = [0,1)$ and define $\alpha$ and $\psi$ as in Item $(\mathrm{vi})$ of Proposition \ref{prop.dictionary} with $\psi$ substituted for $\beta$.\\
\\
Recall from Item $(\mathrm{i})$ of Proposition \ref{prop.dictionary} that $\mathbb{O}_{\psi,2} = \mathbb{O}_{\alpha,2} = I$ where we write $I$ for the identity operator on $L^2([0,1))$. We will construct two functions $f,g:[0,1) \to \mathbb{C}$ satisfying \[ \mathbb{O}_{\alpha,0}[f] = \mathbb{O}_{\alpha,0}[g] = \mathbb{O}_{\psi,0}[f] = \mathbb{O}_{\psi,0}[g] = 0 \] and make the following identifications, where we write $\mathbb{O}_\alpha$ for $\mathbb{O}_{\alpha,1}$ and $\mathbb{O}_\psi$ for $\mathbb{O}_{\psi,1}$.  

\begin{equation} \label{eq.identify} \bigl( f, g,\mathbb{O}_\alpha, I - \mathbb{O}_\alpha, \mathbb{O}_\psi,I - \mathbb{O}_\psi  \bigr)  \mapsto (\imath, \jmath, q_\kappa,\hat{q}_\kappa,q_\lambda,\hat{q}_\lambda) \end{equation}

In order to recover the the result of the calculations in Section \ref{seg.compute} we need the analog of Proposition \ref{prop.CHSH-elementary} for both $\kappa$ and $\lambda$. More explicitly, let $(\gamma,\nu) \in \{(\alpha,\kappa),(\psi,\lambda)\}$ and consider the following equations. \begin{align} \cos^2(\nu) &= \int_0^1 |\mathbb{O}_\gamma[f]|(x)|^2 \deee x  \label{eq.long-a1} \\ & = \int_0^1 |g(x)- \mathbb{O}_\gamma[g](x)|^2 \deee x  \label{eq.long-a2} \\ \sin^2(\nu) & = \int_0^1 |\mathbb{O}_{\gamma}[g](x) |^2 \deee x \label{eq.long-a3}  \\ & =  \int_0^` |f(x)- \mathbb{O}_{\gamma}[f](x) |^2 \deee x  \label{eq.long-a4} \\ \sin(\nu)\cos(\nu) & = \int_0^1  \mathbb{O}_{\gamma}[f](x) \ov{\mathbb{O}_{\gamma}[g](x)} \deee x \label{eq.long-a5} \\ & = -  \int_0^1 \bigl(f(x)-\mathbb{O}_{\gamma}[f](x)\bigr)\bigl(\ov{g(x)}-\ov{\mathbb{O}_{\gamma}[g](x)} \bigr)\deee x \label{eq.long-a6}  \end{align} 
 
 Supposing we have verified (\ref{eq.long-a1}) - (\ref{eq.long-a6}) for both $(\alpha,\kappa)$ and $(\psi,\lambda)$, we now explain how we will be able to conclude that the identification in (\ref{eq.identify}) allows the complete reproduction of the angular CHSH strategy as a statistical spatial strategy. Perform the above construction for the pair $(\kappa,\lambda) = \bigl(\frac{\pi}{3},0\bigr)$ to obtain $f,g:[0,1) \to \mathbb{C}$. Then perform the above construction again for the pair $(\kappa,\lambda) = \bigl(\frac{\pi}{6},-\frac{\pi}{6}\bigr)$ to obtain functions $h,k:[0,1) \to \mathbb{C}$. Define an entangled wavefunction $\Delta:[0,1)^2 \to \mathbb{C}$ by \[ \Delta(s,t) = \frac{f(s)h(t) + g(s)k(t)}{\sqrt{2}} \] The remainder of the realization of the quantum spatial strategy as a statistical spatial strategy can be carried out based on the material from Sections \ref{sec.def} and \ref{sec.CHSH}. The statistical entanglement is captured by the fact that $\Delta$ is not well approximated by functions which split as products.
 
 \subsubsection{Execution of the construction} \label{subsec.execute}
 
  We now perform the required construction. Let $\kappa \in \left \{ - \frac{\pi}{6}, 0\right\}$ and let $\lambda = \kappa + \frac{\pi}{3}$. Define a two dimensional Hilbert space $\mathscr{X} \leq  L^2([0,1))$ as those functions which are constant between the points $\left\{0,\frac{1}{3},\frac{2}{3},1 \right\}$ and have integral $0$. Define $v:[0,1) \to \mathbb{C}$ by \[ v(x) = \begin{dcases} \frac{1}{\sqrt{2}} & \mbox{ if }0 \leq x < \frac{2}{3} \\ -\sqrt{2} & \mbox{ if } \frac{2}{3} \leq x < 1 \end{dcases} \] and $w:[0,1) \to \mathbb{C} $ by \[ w(x) = \begin{dcases} -\sqrt{2} & \mbox{ if }0 \leq x < \frac{1}{3} \\ \frac{1}{\sqrt{2}} & \mbox{ if } \frac{1}{3} \leq j < 1 \end{dcases} \]

  We compute \begin{align*} 1 &= \int_0^1 |v(x)|^2 \deee x = \int_0^1 |w(x)|^2 \deee x \\ 0 & = \int_0^1 v(x) \deee x = \int_0^1 w(x) \deee x \end{align*} so that $v$ and $w$ are elements of $\mathscr{X}$ and $||v|| = ||w|| = 1$. We also find \[ \langle v,w \rangle =\int_0^1 v(x)w(x) \deee x = \frac{1}{2} \] 
 
 We observe that $v$ is in the range of $\mathbb{O}_\alpha$ and is in the kernel of $\mathbb{O}_{\alpha,0}$. Since the range of $\mathbb{O}_\alpha$ has dimension $2$ we find $\mathbb{O}_\alpha - \mathbb{O}_{\alpha,0}  = \langle \cdot, v \rangle v$. Similarly, we have $\mathbb{O}_\psi - \mathbb{O}_{\psi,0} = \langle \cdot, w \rangle w$. Since $v$ and $w$ are unit vectors in the two dimensional space $\mathscr{X}$ and $\langle v,w \rangle = \cos(\kappa - \lambda)$ we may find orthonormal vectors $f,g \in \mathscr{X}$ with $v = \cos(\kappa)f+\sin(\kappa)g$ and $w = \cos(\lambda)f+\sin(\lambda)g$.
 
  \subsubsection{Verification of the construction}

  Define vectors $\hat{v} = -\sin(\kappa)f + \cos(\kappa)g$ and $\hat{w} = \sin(\lambda)f -\cos(\lambda)g$. Since $f$ and $g$ are orthonormal, we see that the pairs $\{v,\hat{v}\}$ and $\{w,\hat{w}\}$ are each an orthonormal basis for $\mathscr{X}$. Therefore if we write $I$ for the identity operator on $\mathscr{X}$ we have $I - \mathbb{O}_\alpha = \langle \cdot, \hat{v} \rangle \hat{v}$ and $I - \mathbb{O}_\psi = \langle \cdot,\hat{w}\rangle \hat{w}$. We make the following observations \begin{align}  \cos(\kappa) & = \langle f ,v \rangle   \label{eq.o-0}\\ & = \langle g, \hat{v} \rangle \label{eq.fuss}  \\ \sin(\kappa) &=  \langle g,v \rangle  \label{eq.o-3} \\ & =-  \langle f, \hat{v} \rangle   \label{eq.o-4} \\ \cos(\lambda) &=  \langle f,w\rangle \label{eq.o-2} \\  &=  \langle g,\hat{w} \rangle \label{eq.o-5} \\ \sin(\lambda) & =  \langle g, w  \rangle  \label{eq.o-2.5}   \\  & =- \langle f, \hat{w} \rangle \label{eq.o-4.5}    \end{align}
  
  We now verify each of the requirements from Section \ref{sec.requirements}, making the following identifications of operators on $\mathscr{Y}$. \[\mathbb{O}_\alpha \mapsto \langle \cdot, v \rangle v \quad \quad  I - \mathbb{O}_\alpha \mapsto \langle \cdot \hat{v}\rangle \hat{v} \quad\quad  \mathbb{O}_\psi \mapsto \langle \cdot, w \rangle w  \quad \quad I - \mathbb{O}_{\psi} \mapsto \langle \cdot, \hat{w} \rangle \hat{w}\] Using these we can set up the following identifications. 
  
  \begin{multicols}{2} \begin{itemize} \item (\ref{eq.long-a1}) for $\kappa$ $\mapsto$ (\ref{eq.o-0}$)^2$ \item (\ref{eq.long-a2}) for $\kappa$ $\mapsto$ (\ref{eq.fuss}$)^2$ \item (\ref{eq.long-a3}) for $\kappa$ $\mapsto$ (\ref{eq.o-3}$)^2$ \item (\ref{eq.long-a4}) for $\kappa$ $\mapsto$ (\ref{eq.o-4}$)^2$ \item (\ref{eq.long-a5}) for $\kappa$ $\mapsto$ (\ref{eq.o-0})(\ref{eq.o-3}) \item (\ref{eq.long-a6}) for $\kappa$ $\mapsto$ (\ref{eq.fuss})(\ref{eq.o-4}) \end{itemize} \columnbreak \begin{itemize} \item (\ref{eq.long-a1}) for $\lambda$ $\mapsto$ (\ref{eq.o-2}$)^2$ \item (\ref{eq.long-a2}) for $\lambda$ $\mapsto$ (\ref{eq.o-5}$)^2$ \item (\ref{eq.long-a3}) for $\lambda$ $\mapsto$ (\ref{eq.o-2.5}$)^2$ \item (\ref{eq.long-a4}) for $\lambda$ $\mapsto$ (\ref{eq.o-4.5}$)^2$ \item (\ref{eq.long-a5}) for $\lambda$ $\mapsto$ (\ref{eq.o-2})(\ref{eq.o-2.5}) \item (\ref{eq.long-a6}) for $\lambda$ $\mapsto$ (\ref{eq.o-5})(\ref{eq.o-4.5})  \end{itemize} \end{multicols} This completes the verification of the construction.
   
   \subsubsection{Interpretation of the construction}
   
   The construction in Section \ref{subsec.execute} can be interpreted as follows. The underlying Hilbert space of the systems consists of functions defined on the unit square which are constant on each of the nine subsquares whose vertices lie in $\frac{1}{3}\mathbb{Z} \times \frac{1}{3}\mathbb{Z}$. We may identify such a function with a function $\psi:\{0,1,2\}^2 \to \mathbb{C}$. For a subset $C$ of $\{0,1,2\}$ we write $\ov{C}$ for $\{0,1,2\} \setminus C$. \\
   \\
   In any round of the game, Alice and Bob may choose subsets $C$ and $D$ of $\{0,1,2\}$. Then they are permitted to make observations corresponding to sums of this function over the respective partitions \[ \{C \times \{0,1,2\},\ov{C} \times \{0,1,2\}\} \quad \{\{0,1,2\} \times D, \{0,1,2\} \times \ov{D}\} \]  In the CHSH game strategy we describe, when Alice and Bob receive the bits $(x,y)$ from the referee they choose $C = \{2x\}$ and $D = \{2y\}$. There is a visualization of this process below in the case $(x,y) = (0,0)$.
   
   \newpage
   
   \begin{center}
   \begin{multicols}{2}

   \begin{tikzpicture}[scale=0.75, transform shape] \node at (1.5,1.5) {$|\psi(0,0)|^2$};
\node at (4.5,1.5) {$|\psi(1,0)|^2$};
\node at (7.5,1.5) {$|\psi(2,0)|^2$};
\node at (1.5,4.5) {$|\psi(0,1)|^2$};
\node at (4.5,4.5) {$|\psi(1,1)|^2$};
\node at (7.5,4.5) {$|\psi(2,1)|^2$};
\node at (1.5,7.5) {$|\psi(0,2)|^2$};
\node at (4.5,7.5) {$|\psi(1,2)|^2$};
\node at (7.5,7.5) {$|\psi(2,2)|^2$};

  \draw[black, ultra thick] (0,0)--(9,0);\draw[black,ultra thick] (0,0)--(0,9);\draw[black,ultra thick] (0,9)--(9,9);\draw[black,ultra thick] (9,0)--(9,9); \draw[black] (3,0)--(3,9);\draw[black] (0,3)--(9,3); \node [below=0.5cm, align=flush center,text width=8cm] at (4.5,0) {Partition for $(x,y) = (0,0)$};\end{tikzpicture}

   \begin{tikzpicture}[scale=0.75, transform shape] 
   
   \node at(1.5,6.5) {Alice's prob.}; \node at (1.5,6) { of answering $0$:};
   \node at (1.5,5) {$|\psi(0,2)|^2$};
   
   \node at (1.5,4.5) {$+|\psi(0,1)|^2$};
   
   \node at (1.5,4) {$+|\psi(0,0)|^2$};
   
   \node at (6,6.25) {Alice's probability of answering $1$:};
   
   \node at (6,5) {$|\psi(1,2)|^2+|\psi(2,2)|^2$};
   
   \node at (6,4.5) {$+|\psi(1,1)|^2+|\psi(2,1)|^2$};
   
   \node at (6,4) {$+|\psi(0,1)|^2+|\psi(0,2)|^2$};

  \draw[black, ultra thick] (0,0)--(9,0);\draw[black,ultra thick] (0,0)--(0,9);\draw[black,ultra thick] (0,9)--(9,9);\draw[black,ultra thick] (9,0)--(9,9);\draw[black] (3,0)--(3,9); \node [below=0.5cm, align=flush center,text width=8cm] at (4.5,0) {Alice sums vertically};\end{tikzpicture}

   \begin{tikzpicture}[scale=0.75, transform shape] \node at (1.5,1.5) {$|\psi(0,0)|^2$};
   
   \node at (6,1.5) {$|\psi(0,1)|^2+|\psi(0,2)|^2$};

   \node at (1.5,6.4) {$|\psi(0,2)|^2$}; \node at (1.5,5.6) {$+|\psi(0,1)|^2$};

   \node at (6,6.4) {$|\psi(1,2)|^2+|\psi(2,2)|^2$}; \node at (6,5.6) {$+|\psi(1,1)|^2+|\psi(2,1)|^2$};

  \draw[black, ultra thick] (0,0)--(9,0);\draw[black,ultra thick] (0,0)--(0,9);\draw[black,ultra thick] (0,9)--(9,9);\draw[black,ultra thick] (9,0)--(9,9); \draw[black] (3,0)--(3,9);\draw[black] (0,3)--(9,3); \node [below=0.5cm, align=flush center,text width=8cm] at (4.5,0) {Partially collapsed function for $(x,y) = (0,0)$};\end{tikzpicture} \\ \vspace{1 cm}

   \begin{tikzpicture}[scale=0.75, transform shape] 
   
   \node at(4.5,2) {Bob's probability of answering $0$:};
   \node at (4.5,1) {$|\psi(0,0)|^2+|\psi(0,1)|^2+|\psi(0,2)|^2$};

   \node at(4.5,7) {Bob's probability of answering $1$:};
   \node at (4.5,6) {$|\psi(0,2)|^2+\psi(1,2)|^2+|\psi(2,2)|^2$};

   \node at (4.5,5) {$+|\psi(0,1)|^2+|\psi(1,1)|^2+|\psi(1,2)|^2$};

  \draw[black, ultra thick] (0,0)--(9,0);\draw[black,ultra thick] (0,0)--(0,9);\draw[black,ultra thick] (0,9)--(9,9);\draw[black,ultra thick] (9,0)--(9,9);\draw[black] (0,3)--(9,3); \node [below=0.5cm, align=flush center,text width=8cm] at (4.5,0) {Bob sums horizontally};\end{tikzpicture}
  
  \end{multicols}
  
 Figure 1: Alice and Bob's measurement procedures when $(x,y) = (0,0)$.

\vspace{0.5 cm}

   \end{center}
  
  The idea is that for $\kappa \in [-\pi,\frac{2\pi}{3}]$ we may uniquely define a function $v_\kappa,w_\kappa:\{0,1,2\} \to \mathbb{C}$ as the solution to the following overdetermined system of equations.
  
\begin{align*} \left(-\sqrt{2},\frac{1}{\sqrt{2}},\frac{1}{\sqrt{2}}\right )& = \cos(\kappa)(v_\kappa(0),v_\kappa(1),v_\kappa(2)) + \sin(\kappa)(w_\kappa(0),w_\kappa(1),w_\kappa(2))   \\ \left(\frac{1}{\sqrt{2}},\frac{1}{\sqrt{2}},-\sqrt{2}\right )  & =\cos\left(\kappa+\frac{\pi}{3}\right)(v_\kappa(0),v_\kappa(1),v_\kappa(2))  + \sin\left(\kappa + \frac{\pi}{3}\right)(w_\kappa(0),w_\kappa(1),w_\kappa(2)) \\1&=  \frac{1}{3}\left( |v_\kappa(0)|^2+|v_\kappa(1)|^2+|v_\kappa(2)|^2)  \right)=\frac{1}{3}\left( |w_\kappa(0)|^2+|w_\kappa(1)|^2+|w_\kappa(2)|^2)  \right) \\ 0&=v_\kappa(0)+v_\kappa(1)+v_\kappa(2)=w_\kappa(0)+w_\kappa(1)+w_\kappa(2)\end{align*}
  
  If we define $\Delta:\{0,1,2\} \to \mathbb{C}$ be given by \[ \Delta(j,k) = \frac{v_0(j)v_{-\frac{\pi}{6}}(k)+w_0(j)w_{-\frac{\pi}{6}}(k)}{\sqrt{2}} \] then the procedure described in Figure 1 produces the probabilities required for the CHSH strategy as in (\ref{eq.angles}). In physical terms, we may visualize the two orthogonal axes in Figure 1 as measurement apparatuses belonging to Alice and Bob. If Alice receives the bit $x$, she measures the probability that she observes the particle to the left of a divider placed in her axis at $\frac{x+1}{3}$ and uses this as her probability of responding with the answer $0$. Similarly, if Bob receives the bit $y$ he measures the probability that he observes the particle below the divider placed in his axis at $\frac{x+1}{3}$ and uses this as his probability of answering $0$. The entangled nature of the wavefunction $\Delta$ allows for a higher value of the game than if Alice and Bob were performing the same process for particles on their individual axes.

  \section{Proof of Theorem \ref{thm.main}} \label{sec.main}
  
Throughout Section \ref{sec.main} we fix a nonlocal game $\mathfrak{G} = (\mathscr{X},\mathscr{Y},\mathscr{A},\mathscr{B},\pi,D)$.  It is transparent from the definitions that the strategy spaces for $\mathfrak{G}$ depend only on the question and answer sets, so the question distribution and payoff function are not relevant to establishing Theorem \ref{thm.main}. Write $n = |\mathscr{A}|$ and $m = |\mathscr{B}|$ and identify $(\mathscr{A},\mathscr{B})$ with $(\mathbb{Z}_n,\mathbb{Z}_m)$. We will use natural numbers in the least residue system to refer to the elements of these groups. For a function $\phi:\mathbb{Z}_n \times \mathbb{Z}_m \to \mathbb{C}$ let $\hat{\phi}:\mathbb{Z}_n \times \mathbb{Z}_m \to \mathbb{C}$ be the Fourier transform given by \[ \hat{\phi}(j,k) = \frac{1}{nm} \sum_{(s,t) \in \mathbb{Z}_n \times \mathbb{Z}_m} e_n(-js)e_m(-kt)\phi(s,t) \] Define two countable discrete groups  \begin{equation} \label{eq.free} G_A = \freeprod {\,}_{x \in \mathscr{X}}\,\, \mathbb{Z}_n \quad \mbox{ and } \quad G_B = \freeprod {\,}_{y \in \mathscr{Y}} \,\,\mathbb{Z}_m \end{equation}  where the large asterisks denote free products.

 \subsection{Observation/dynamic duality, quantum case} \label{subsec.staticq}
  
In Section \ref{subsec.staticq} we define dynamical or dual analogs of the quantum observation objects discussed in Section  \ref{sec.def}.

\begin{definition} For a bare strategy $(p_{x,y})_{(x,y) \in \mathscr{X} \times \mathscr{Y}}$ define the \textbf{dual game} value at $p$ by \[ \hat{G}(p) =  \sum_{(x,y) \in \mathscr{X} \times \mathscr{Y}} \pi(x,y) \sum_{(j,k) \in \mathbb{Z}_n \times \mathbb{Z}_m} \hat{D}(x,y,j,k) p_{x,y}(j,k) \] By unitarity of the Fourier tranform we have \[ \sum_{(j,k) \in \mathbb{Z}_n \times \mathbb{Z}_m} \hat{D}(x,y,j,k)p_{x,y}(j,k) = \sum_{(j,k) \in \mathbb{Z}_n \times \mathbb{Z}_m} D(x,y,j,k)\hat{p_{x,y}}(j,k) \] for all $(x,y) \in \mathscr{X} \times \mathscr{Y}$. Therefore $\hat{\mathfrak{G}}(p) = \mathfrak{G}(\hat{p})$. \end{definition}

We note the Fourier transform of a pointwise positive function is a so-called positive definite function, and these are the natural payoff functions for the duals of nonlocal games. We now define the dynamical analog of a projection valued measure. 

\begin{definition} A \textbf{wheel} on a Hilbert space $H$ is a unitary representation of $\mathbb{Z}_n$ on $H$.  \end{definition}

\begin{remark} \label{rem.wheel} Let $A_1,\ldots,A_n$ be a projection valued measure on $H$. Then $A_1,\ldots,A_n$ gives rise to a wheel $u(0),\ldots,u(n-1)$ by letting \begin{equation} u(k) = \sum_{j=0}^{n-1} e_n(kj)A_j \label{eq.wheel} \end{equation}  \end{remark}

\begin{remark} Let $A_1,\ldots,A_n$ and $B_1,\ldots,B_m$ be two commuting projection valued measures on $H$ giving rise to two commuting wheels $u(0),\ldots,u(n-1)$ and $v(0),\ldots,v(m-1)$. For a unit vector $\psi \in H$ we observe \[ \langle u(j)\psi ,\,v(k) \psi \rangle = \sum_{(s,t) \in \mathbb{Z}_n \times \mathbb{Z}_m} e_n(-js)e_m(-kt)\langle A_t \psi ,\, B_s \psi \rangle \] and so if we write $p(j,k) = \frac{1}{nm} \langle u(j)\psi,\, v(k)\psi \rangle$ and $q(j,k) = \langle A_j \psi, \,B_k \psi \rangle$ then we have $p = \hat{q}$ \label{rem.fourier} \end{remark}

\begin{definition} We have the following dual versions of the objects from Definition \ref{def.quantum} \begin{itemize} \item We define a bare strategy $(p_{x,y})_{(x,y) \in \mathscr{X} \times \mathscr{Y}}$ for $\mathfrak{G}$ to be \textbf{unitary commuting strategy} if it is generated as follows. Consider a  Hilbert space $L$ and assume we have a unitary representation $\underline{a}$ of $G_A$ on $L$ and a unitary representation $\underline{b}$ of $G_B$ on $L$ which commutes with $\underline{a}$. Then we set \begin{equation} \label{eq.qucom} p_{x,y}(a_j,b_k) = \frac{1}{nm} \langle \underline{a}(x,j)\psi ,\,  \underline{b}(y,k) \psi \rangle \end{equation} for some unit vector $\psi \in L$. We define the \textbf{unitary commuting strategy space} of $\mathfrak{G}$ by \[ \mathsf{UCo}(\mathfrak{G}) =\Bigl  \{ \mathbf{p} \in [0,1]^{\mathscr{X} \times \mathscr{Y} \times \mathscr{A} \times \mathscr{B}} : \mathbf{p} \mbox{ is a unitary commuting strategy for }\mathfrak{G} \Bigr \}  \]  \item We define a unitary commuting strategy $(p_{x,y})_{(x,y) \in \mathscr{X} \times \mathscr{Y}}$ for $\mathfrak{G}$ to be a \textbf{unitary spatial strategy} if it is generated as follows. Consider  Hilbert spaces $H$ and $K$ and let $\underline{a}$ be a unitary representation of $G_A$ on $H$ and let $\underline{b}$ be a unitary representation of $G_B$ on $K$. Then we set \begin{equation} \label{eq.spatial-3} p_{x,y}(a_j,b_k) = \frac{1}{nm} \langle (\underline{a}(x,j)  \otimes \underline{b}(y,k)) \psi, \, \psi \rangle \end{equation} for some unit vector $\psi \in H \otimes K$. We define the \textbf{unitary spatial strategy space} of $\mathfrak{G}$ by \[ \mathsf{USp}(\mathfrak{G}) =\Bigl  \{ \mathbf{p} \in [0,1]^{\mathscr{X} \times \mathscr{Y} \times \mathscr{A} \times \mathscr{B}} : \mathbf{p} \mbox{ is a unitary spatial strategy for }\mathfrak{G} \Bigr \}  \]  \end{itemize} \end{definition}

 \begin{remark} From Remark \ref{rem.fourier} we see that \begin{equation} \label{eq.amanda-4} \mathsf{UCo}(\mathfrak{G}) = \Bigl \{ (\hat{p}_{x,y})_{(x,y) \in \mathscr{X} \times \mathscr{Y}}: (p_{x,y})_{(x,y) \in \mathscr{X} \times \mathscr{Y}} \in \mathsf{QCo}(\mathfrak{G}) \Bigr \} \end{equation} and \begin{equation} \label{eq.amanda-5} \mathsf{USp}(\mathfrak{G}) = \Bigl \{ (\hat{p}_{x,y})_{(x,y) \in \mathscr{X} \times \mathscr{Y}}: (p_{x,y})_{(x,y) \in \mathscr{X} \times \mathscr{Y}} \in \mathsf{QSp}(\mathfrak{G}) \Bigr \} \end{equation}  \end{remark}

\subsection{Observation/dynamic duality, statistical case} \label{subsec.statlate}

In Section \ref{subsec.statlate} we define ergodic or dual versions of the statistical observation objects from \ref{subsec.measstrat}. We will use the theory of measure preserving actions of countable discrete groups on standard probability spaces as developed in \cite{MR2583950}.  If $(\Omega,\mu)$ is a standard probability space and $n \in \mathbb{N}$ we write $\mathrm{Aut}(\Omega,\mu)$ for the group of measure preserving transformations of $\Omega$. We write $\mathrm{Aut}(\Omega,\mu)_n$ for the set of all transformations $T \in \mathrm{Aut}(\Omega,\mu)$ such that all orbits have size $n$. We also fix an ambient background total linear $\leq$ order on $\Omega$. \\
\\
Given an observable $\alpha$ of resolution $n$ on $\Omega$ we can define an associated $T \in \mathrm{Aut}(\Omega,\mu)_n$ by letting $T$ rotate the orbits by one modulo $n$, where this notation is defined according to $c_\alpha$. If $T \in \mathrm{Aut}(\Omega,\mu)_n$ we can construct an observable $\alpha$ on $\Omega$ by taking $\sim_\alpha$ to be the orbit equivalence relation of $T$ and taking $c_\alpha$ according to the restriction of $\leq$ to each $T$-orbit. In this case we write $c_T$ for $c_\alpha$. Let $\delta^k_T:\Omega \to \Omega$ be the map which sends each point to the $k^{\mathrm{th}}$ element in its $T$-orbit and for $f: \Omega \to \mathbb{C}$ and $s \in \Omega$ define $f^T_s:\mathbb{Z}_n \to \mathbb{C}$ by $f_s^T(k) = f(\delta^k_T(s))$. We take the superscript index on $\delta^k_T$ and the argument of $f_s^T$ modulo $n$. Then we have that \begin{equation} \label{eq.expect} \mathbb{O}_{\alpha,k}[f](s) = \begin{dcases} f(s) & \mbox{ if } 0 \leq c_T(s) \leq k \\  \frac{f_s^T(k+1)+\cdots+f_s^T(n)}{n-k+1} & \mbox{ if } k < c_T(s) \leq n-1 \end{dcases} \end{equation}

\begin{definition} For $T \in \mathrm{Aut}(\Omega,\mu)_n$ we define the $T$\textbf{-local Fourier transform} for $f \in L^2(\Omega,\mu)$ to be the function $\mathcal{F}_T[f] \in  L^2(\Omega,\mu)$ given by \[ \mathcal{F}_T[f](\delta^j_T(s)) = \frac{1}{n} \sum_{k=0}^{n-1} e_n(-kj)f_s^T(k) \] \end{definition}

For $T \in \mathrm{Aut}(\Omega,\mu)$ define the Koopman operator $\kappa_T$ on $L^2(\Omega,\mu)$ by letting $\kappa_T[f](s) = f(T^{-1}s)$. Even when $L^2(\Omega,\mu)$ is infinite dimensional, the eigenvalues of $\kappa_T$ are the $n^{\mathrm{th}}$ roots of unity and the eigenspace of $e_n(-j)$ is given by all functions of the form $f(s) = q(s)e_n(-jc_T(s))$ where $q(s):\Omega \to \mathbb{C}$ is a $T$-invariant function in $L^2(\Omega,\mu)$ . If we write $A_j$ for the projection onto this eigenspace then because $A_j$ commutes with $\kappa_T$ and has a one-dimensional range inside each $\kappa_T$-invariant subspace, we have $A_jf$ can be computed locally as the projection onto the subspace of the $\sim_\alpha$-class spanned by $e(jc_T(\cdot))$. More explicitly, we have \[ A_j[f](s) =  \left( \frac{1}{n} \sum_{k=0}^{n-1} e_n(-kj)f_s^T(k)  \right) e_n(-jc_T(s))= \mathcal{F}_T[f](\delta^j_T(s)) \cdot e_n(-jc_T(s)) \] Using this we see that the formulas $A_0+\cdots+A_{n-1} = I$ and \[ \sum_{j=0}^{n-1} e_n(-j)A_j = \kappa_T\] become cases of the $T$-local Fourier inversion formulas for $\mathbb{Z}_n$. Moreover, if we write $A_\ast$ for the orthogonal projection onto the vector $1+e_n(-1)+e_n(-2)+\cdots+e_n(-n+1)$ then $k \in \{-1,\ldots,n-2\}$ we have that $A_{\ast} +A_0+ \cdots +A_k$ is the orthogonal projection onto the span of the functions \[ 1,e_n(-1),e_n(-2),\ldots,e_n(-k),e_n(-k-1)+\cdots+e(-n+1) \] Therefore we see that $\mathcal{F}_T(A_{\ast}+A_0+\cdots+A_k)$ is the orthogonal projection onto the span of the functions \[ \mathbf{1}_{\delta^0_T}, \ldots, \mathbf{1}_{\delta^{-k}_T}, \mathbf{1}_{\delta^{-k-1}_T}+\cdots+\mathbf{1}_{\delta^{-n+1}_T} \] where $\mathbf{1}_{\delta^\ell_T}$ represents the indicator function of $\delta^\ell_T(s)$. Thus we can see from (\ref{eq.expect}) that $\mathcal{F}_T(A_{\ast}+A_0+\cdots+A_k) = \mathbb{O}_{\alpha,k}$ and so the relationship described in Remark \ref{rem.wheel} applies to the projection valued measure \[ \{\mathbb{O}_{\alpha,k} - \mathbb{O}_{\alpha,k-1}:0 \leq k \leq n-1\} \] on $L^2(\Omega,\mu)$ and the wheel $\kappa_T:\mathbb{Z}_n \to  \mathrm{U}(L^2(\Omega,\mu))$.

\begin{remark} Let $\alpha$ and $\beta$ be two consistent observables. Then $\alpha$ and $\beta$ give rise to two commuting transformations $T \in \mathrm{Aut}(\Omega,\mu)_n$ and $S \in \mathrm{Aut}(\Omega,\mu)_m$, which in turn give rise to two commuting wheels $\kappa_T$ and $\kappa_S$. Let $f \in L^2(\Omega,\mu)$ with \[ \int_\Omega |f(u)|^2 \deee \mu(u)=1 \] and define \begin{align*}  p(k,j)& = \frac{1}{nm} \int_\Omega f(T^{-k}u)f(S^{-j}u) \deee \mu(u) \\ q(k,j)& =  \int_\Omega \bigl(\mathbb{O}_{\alpha,k}[f](u)- \mathbb{O}_{\alpha,k-1}[f](u)\bigr)\bigl( \ov{\mathbb{O}_{\beta,j}[f](u)} - \ov{\mathbb{O}_{\beta,j-1}[f](u)}\bigr) \deee \mu(u) \end{align*} Then we have $p = \hat{q}$ by Remark \ref{rem.fourier} \label{rem.fourier-2} \end{remark}

\begin{definition} \label{def.measstrat-2} We have the following dual versions of the objects from Definition \ref{def.measstrat}. \begin{itemize}\item We define a bare strategy for $\mathfrak{G}$ to be an \textbf{ergodic commuting strategy} if it is generated by the following data. \begin{itemize} \item A standard probability space $\Omega$ which represents a common sample space belonging to both Alice and Bob. \item For each $x \in \mathscr{X}$ let $T_x \in \mathrm{Aut}(\Omega,\mu)_n$ be a transformation belonging to Alice. \item For each $y \in \mathscr{Y}$ let $S_y \in \mathrm{Aut}(\Omega,\mu)_m$ be a transformation belonging to Bob.  \item We stipulate that for each pair $(x,y) \in \mathscr{X} \times \mathscr{Y}$ we have that $T_x$ commutes with $S_y$. \item A function $f:\Omega \to \mathbb{C}$ with \[ \int_\Omega |f(u)|^2 \deee \mu(u) = 1 \]  \end{itemize} Given these data we set \begin{equation} p_{x,y}(a_k,b_j) = \frac{1}{nm} \int_\Omega f(T_x^{-k}u)\ov{f(S_y^{-j}u) }\deee \mu(u)  \label{eq.excom} \end{equation} We define the \textbf{ergodic commuting strategy space} of $\mathfrak{G}$ by \[ \mathsf{ErgCo}(\mathfrak{G}) =\Bigl  \{ \mathbf{p} \in [0,1]^{\mathscr{X} \times \mathscr{Y} \times \mathscr{A} \times \mathscr{B}} : \mathbf{p} \mbox{ is an ergodic commuting strategy for }\mathfrak{G} \Bigr \}  \]  \item We define an ergodic commuting strategy for $\mathfrak{G}$ to be a \textbf{ergodic spatial strategy} if it is generated by the following data. \begin{itemize} \item Standard probability spaces $(\Lambda,\nu)$ and $(\Pi,\eta)$ which represent two spatially separated sample spaces belonging to Alice and Bob respectively.  \item For each $x \in \mathscr{X}$ let $T_x \in \mathrm{Aut}(\Lambda,\nu)_n$ be a transformation belonging to Alice.   \item For each $y \in \mathscr{Y}$ let $S_y \in \mathrm{Aut}(\Pi,\eta)_m$ be a transformation belonging to Bob. \item A function $f: \Lambda \times \Pi \to \mathbb{C}$ with \[ \int_{\Lambda \times \Pi} |f(s,t)|^2 \deee(\nu \times \eta)(s,t) = 1 \] \end{itemize} Given these data, we set \begin{equation} p_{x,y}(a_k,b_j) = \frac{1}{nm} \int_{\Lambda \times \Pi} f(T^{-k}_xs,t)\ov{f(s,S^{-j}_yt)} \deee (\nu \times \eta)(s,t) \end{equation} We define the \textbf{ergodic spatial strategy space} of $\mathfrak{G}$ by \[ \mathsf{ErgSp}(\mathfrak{G}) = \Bigl \{\mathbf{p} \in [0,1]^{\mathscr{X} \times \mathscr{Y} \times \mathscr{A} \times \mathscr{B}}: \mathbf{p} \mbox{ is a ergodic spatial strategy for }\mathfrak{G} \Bigr \} \] \end{itemize}  \end{definition}

 \begin{remark} From Remark \ref{rem.fourier-2} we see that \begin{equation} \label{eq.amanda-2} \mathsf{ErgCo}(\mathfrak{G}) = \Bigl \{ (\hat{p}_{x,y})_{(x,y) \in \mathscr{X} \times \mathscr{Y}}: (p_{x,y})_{(x,y) \in \mathscr{X} \times \mathscr{Y}} \in \mathsf{StatCo}(\mathfrak{G}) \Bigr \} \end{equation} and \begin{equation} \label{eq.amanda-3} \mathsf{ErgSp}(\mathfrak{G}) = \Bigl \{ (\hat{p}_{x,y})_{(x,y) \in \mathscr{X} \times \mathscr{Y}}: (p_{x,y})_{(x,y) \in \mathscr{X} \times \mathscr{Y}} \in \mathsf{StatSp}(\mathfrak{G}) \Bigr \} \end{equation}  \end{remark}

\subsection{Gaussian Hilbert spaces in representation theory and ergodic theory} \label{sec.alex}

By Appendix A of \cite{MR2583950} we may assume that we are dealing with real Hilbert spaces. The analysis in Section \ref{sec.alex} reflects the analysis in Appendix H of the same reference and also the main topic of \cite{MR1474726}.  If $F \subseteq G$ is finite we let $\imath_F:\mathbb{R}^G \to \mathbb{R}^F$ be the canonical projection map. \\
\\
Let $\mathbf{p}$ be a unitary commuting strategy given by a pair of commuting representations $\underline{a}:G_A \to \mathrm{U}(H)$ and $\underline{b}:G_B \to \mathrm{U}(H)$. Let $p:G \to \mathbb{C}$ be the positive definite function defined by \[ p(g,h) = \langle \underline{a}(g) \psi, \, \underline{b}(h) \psi \rangle \] for $g \in G_A$ and $h \in G_B$. If $F \subseteq G$ is finite we define $p_F$ to be the $F \times F$ submatrix of $p$.  We define the Gaussian probability measure $\gamma_\mathbf{p}$ on $\mathbb{R}^G$ with the product topology by considering a bounded Borel function $\phi:\mathbb{R}^F \to \mathbb{C}$ and setting \begin{equation} \label{eq.favorite}  \int_{\mathbb{R}^G} \phi(\imath_F(z)) \deee \gamma_\mathbf{p}(z)  = \frac{1}{\sqrt{(2\pi)^{|F|} \,\mathrm{det}\, p_F} }\int_{\mathbb{R}^F} \phi(y) \exp\left( - \frac{y^\ast p_F^{-1}y}{2} \right) \deee y \end{equation}

There is a minor technicality if $p_F$ is singular for some $F$, but we can avoid it by considering a measure supported on the closure of the subspace of $\mathbb{R}^G$ generated by the subspaces $\imath_F^{-1}(\sqrt{p_F} (\mathbb{R}^F))$ for finite subsets $F \subseteq G$. In this case the determinant in the normalizing prefactor in (\ref{eq.favorite}) should be replaced with the the product of the positive eigenvalues of $p_F$. Note that the Kolmogorov consistency theorem implies that specifying the finite dimensional marginals of $\gamma_\mathbf{p}$ as in (\ref{eq.favorite}) suffices to determine the integral of all bounded Borel functions on $\mathbb{R}^G$. For $g \in G$ write $\imath_g$ for $\imath_{\{g\}}$. Then the basic theory of Gaussian probability measures implies \begin{equation} \label{eq.nicenice} \int_{\mathbb{R}^G} \imath_g(z)\imath_h(z) \deee \gamma_\mathbf{p} (z)  = p(g,h)  \end{equation} We can define the left shift action $B:G \to \mathrm{Aut}(\mathbb{R}^G,\gamma_\mathbf{p})$ by $B_g[z](h) = z(g^{-1}h)$. Then (\ref{eq.nicenice}) becomes \begin{equation} \label{eq.free-2} \int_{\mathbb{R}^G} \imath_{1_G}(B_gz)\imath_{1_G}(B_h z) \deee \gamma_\mathbf{p}(z) \end{equation} Identify $x \in \mathscr{X}$ and $y \in \mathscr{Y}$ with the indexes of the corresponding free products (\ref{eq.free}) and write $(x,k)$ for the one-symbol word consisting of the copy of the number $k$ in the $x$ factor. Adopt a similar notation for $(y,j)$. By specializing (\ref{eq.free-2}) to $g = (x,k)$ and $h = (y,j)$ we can choose $\imath_{1_G} \in L^2(\mathbb{R}^G,\gamma_\mathbf{p})$ as the wavefunction to obtain a realization of $\mathbf{p}$ as an ergodic commuting strategy. Thus we have shown the following.

\begin{lemma} We have $\mathsf{UCo}(\mathfrak{G}) = \mathsf{ErgCo}(\mathfrak{G})$. \label{lem.nicenice} \end{lemma}

Now, let $\mathbf{p}$ be a unitary spatial strategy. We may assume that the representation $\underline{a} \otimes \underline{b}:G_A \times G_B \to \mathrm{U}(H \otimes K)$ which defines $\mathbf{p}$ is irreducible, as this will correspond to a convex combination of strategies. Let $\rho \in H$ and $\vartheta \in K$ be any two unit vectors. Define positive define functions $p_A:G_A \to \mathbb{C}$ and $p_B:G_B \to \mathbb{C}$ by \[ p_A(g,\tau) = \langle \underline{a}(g) \rho,\, \underline{a}(\tau)\rho \rangle \mbox{ and } p_B(h,\delta) = \langle \underline{b}(h)\vartheta,\, \underline{b}(\delta) \vartheta \rangle \] Perform the above Gaussian construction to obtain probability measures $\gamma_A$ on $\mathbb{R}^{G_A}$ and $\gamma_B$ on $\mathbb{R}^{G_B}$ such that \[ \int_{\mathbb{R}^{G_A}} \imath_{g}(z)\ov{\imath_\tau(z)} \deee \gamma_A (z) = p_A(g,\tau) \] \and \[ \int_{\mathbb{R}^{G_B}} \imath_{h}(w)\ov{\imath_\delta(w)} \deee \gamma_B(w) = p_B(h,\delta) \] so therefore  \begin{equation} \label{eq.amanda-1} \int_{\mathbb{R}^{G_A} \times \mathbb{R}^{G_B}} \imath_{g}(z)\ov{\imath_\tau(z)} \imath_{h}(w)\ov{\imath_\delta(w)} \deee(\gamma_A \times \gamma_B)(z,w) = \bigl\langle (\underline{a}(g)\rho) \otimes (\underline{b}(h)\vartheta) \, \big | \, (\underline{a}(\tau)\rho) \otimes (\underline{b}(\delta)\vartheta)  \bigr \rangle  \end{equation} Since we have assumed $\underline{a} \otimes \underline{b}$ is irreducible, we can find a function $\chi:G_A \times G_B \to \mathbb{C}$ such that if $\psi$ is the original wavefunction in $H \otimes K$ then we have \begin{equation} \label{eq.peter-1} \psi = \sum_{(g,h) \in G_A \times G_B} \chi(g,h) \cdot[(\underline{a} \otimes \underline{b})(g,h)](\rho \otimes \vartheta) \end{equation} If we define a wavefunction $f \in L^2(\mathbb{R}^{G_A} \times \mathbb{R}^{G_B},\gamma_A \times \gamma_B)$ by \[ f = \sum_{(g,h) \in G_A \times G_B} \chi(g,h)\imath_g \imath_h \] then (\ref{eq.amanda-1}) and (\ref{eq.peter-1}) together imply that specializing $g$ to $(x,a_k)$ and $h$ to $(y,b_j)$ gives a realization of $\mathbf{p}$ as an ergodic spatial strategy. Thus we have shown the following.

\begin{lemma} We have $\mathsf{USp}(\mathfrak{G}) = \mathsf{ErgSp}(\mathfrak{G})$. \label{lem.nice} \end{lemma}

From (\ref{eq.amanda-5}) and (\ref{eq.amanda-3}) and Lemma \ref{lem.nice} we may conclude that $\mathsf{QSp}(\mathfrak{G}) = \mathsf{StatSp}(\mathfrak{G})$. From (\ref{eq.amanda-4}) and (\ref{eq.amanda-2}) and Lemma \ref{lem.nicenice} we may conclude that $\mathsf{QCo}(\mathfrak{G}) = \mathsf{StatCo}(\mathfrak{G})$. This completes the proof of Theorem \ref{thm.main}.

\section{Conclusions and outlook}

We have described a correspondence between quantum measurement procedures based on applying projection-valued measures to a vector in Hilbert space and statistical measurement procedures based on partitioning a probability space. Given our results that the sets of possible statistical correlations coincide with the sets of possible quantum correlations, it is likely that the statistical measurement procedures we describe are more physically realistic. The concept of a projection valued measure in a Hilbert space is somewhat harder visualize than the concept of a partition of a probability space, especially when the partitions used have an explicit form as with those used in our example of the CHSH game.\\
\\
More broadly, it seems plausible that our ideas could be used to derive appropriate experimental procedures for implementing nonlocal games. Given the game in 'quantum form', the method we describe produces a game with the same value which has a natural description in terms of measuring whether a particle is observed to the left or right of certain dividing points in an interval. It would be interesting to come up with explicit calculations of what this set up looks in well-known cases other than the CHSH game.

\bibliographystyle{plain}
\bibliography{/Users/peterburton/Documents/Math/Bibliography/pjburtonbibliography.bib}

\texttt{burtonpeterj@icloud.com}

\end{document}